# Practical Challenges with Spreadsheet Auditing Tools

Daniel Kulesz, Jan-Peter Ostberg

Institute of Software Technology,
University of Stuttgart, Germany

firstname.lastname@informatik.uni-stuttgart.de

**ABSTRACT**

*Just like other software, spreadsheets can contain significant faults. Static analysis is an accepted and well-established technique in software engineering known for its capability to discover faults. In recent years, a growing number of tool vendors started offering tools that allow casual end-users to run various static analyses on spreadsheets as well.*
*We supervised a study where three undergraduate software engineering students examined a selection of 14 spreadsheet auditing tools, trying to give a concrete recommendation for an industry partner. Reflecting on the study's results, we found that most of these tools do provide useful aids in finding problems in spreadsheets, but we have also spotted several areas where tools had significant issues. Some of these issues could be remedied if spreadsheet auditing tool vendors would pick up some ideas of static analysis tools for traditional software development and adopt some of their solution approaches.*

## 1. INTRODUCTION

Just like other software, spreadsheets are known to contain faults. Spreadsheets are often developed by end-users. Several experiments indicate that spreadsheets are more fault-prone than other software [Powell et al., 2008]. The EuSpRiG has collected evidence of cases where spreadsheet faults caused significant damage. The awareness for these risks has risen in recent years and various new laws like the Sarbanes-Oxley Act 404 are interpreted as demanding more controls and quality assessments on spreadsheets.

It is yet unclear, though, how these risks should be addressed. One reason for this is that it is unclear which spreadsheet characteristics are desirable and which ones should be considered harmful. Research has not provided enough evidence on this yet and practitioners' recommendations are often conflicting with each other [Kulesz, 2011].

In contrast to spreadsheets, there is more agreement about desirable and undesirable characteristics of traditional software backed up by numerous scientific studies [Beck et. al., 1999] [Mäntylä, 2003]. Furthermore, professional software engineers are well aware of the fact that faulty software can result in significant damage and try to address this risk already in early phases of development. One popular approach to do this is by applying static analysis on source code. This can only be feasibly executed with the support of proper tools. In the following, we will refer to these tools used in non-spreadsheet software development as TSATs (Traditional Static Analysis Tools).

Spreadsheets can be seen as programming languages and products like Microsoft Excel as execution environments for them. Hence, static analysis is principally applicable to spreadsheets as well. Despite the uncertainty about desirable characteristics, a growing number of tool vendors is offering tools which allow casual end-users to run static analyses on spreadsheets. According to [Nixon and O'Hara, 2001] these mostly fully-automated tools claim to help in finding faults before they can cause any damage mostly by:

- Providing aids (i.e. a different visualization) which allow the users to better understand the spreadsheet and its internal structure
- Directly identifying potentially faulty cells (by matching them against "smell" patterns)
- Identifying unique formulas in order to allow the auditor to narrow the scope of the inspection

But is this really enough? How do spreadsheet auditing tools find potential faults? How useful are they for casual spreadsheet users? And are there areas where they can learn from the TSATs?

## 2. RELATED WORK

Several years ago, Nixon and O'Hara compared five spreadsheet auditing tools by running them on a single spreadsheet which was seeded with 17 faults of different types [Nixon and O'Hara, 2001]. They compared and graded the detection rates of these tools on a four-level scale. Overall, they came to the conclusion that all evaluated tools were indeed useful, although they all had problems with certain types of faults. It is notable that they treated the auditing functions built into Microsoft Excel also as a tool.

In a later study [Howard, 2007], Howard analyzed 16 tools on the market, putting them in three categories: "Auditor's tools" (5 products), "Control & Compliance tools" (8 products) and "Automation tools" (2 products). According to Howard, sometimes the functionality provided by the five "auditor's tools" was also contained in some of the "control & compliance tools", making this separation a bit fuzzy. In the final conclusion, Howard provides a comparison table which lists various capabilities of each tool (i.e. detecting circular references, detecting logic errors or showing precedents) without actually rating their quality.

The study of [Powell et. al., 2007] did a complete audit of more than 50 spreadsheets with two spreadsheet auditing tools and manual inspections. Although the authors observed a high rate of false positives reported by the tools, they were convinced that this approach was still more effective than manual inspections without tool support. Another prior study by [Clermont et. al., 2002] analyzed only their own spreadsheet auditing tool, but on a selection of 78 spreadsheets from industry. The authors came to the conclusion that the tool was helpful for end-users by increasing the understanding of their spreadsheets and showing them potential faults through irregularities in the visualized patterns. In several other studies, Hermans observed similar effects when running self-developed tools on spreadsheets from the industry and interviewing end-users about the tools' results [Hermans, 2013].

Reflecting on those studies, most of our initial questions still remained open. The two initially mentioned studies reviewed the available tools only from the perspective of "full-time spreadsheet auditors", claiming that the tools were useful for them. But how useful are the tools for casual end-users? Ratings were given context-independent and using spreadsheets with seeded errors. The applicability of these tools to real-world spreadsheets has not been reviewed. This was done in the three latter studies which incorporated real end-users and real spreadsheets, but they only evaluated one resp. two single tools.

We found no study that compared spreadsheet auditing tools with static analysis tools for traditional software, although they share many characteristics.

## 3. STATIC ANALYSIS

Static analysis is comparable to the idea of destruction-free quality checks in the manufacturing industry. The analysis should only report possible defects but not alter the analysed object by any means. Research and development in the field of static analysis has been going on for years now and has produced many highly functional attack vectors to conquer issues related to product quality. In the following, we will take a short look at common approaches of static analysis:

**Data-flow analysis** [Taylor et al. 1980] tracks the possible paths and states the data can take. Its output is a graph which represents every change of a variable in the control flow of a program's structure by a node. Data-flow analysis is especially good at detecting data anomalies, like lost updates or dead code.

**Abstract interpretation** [Cousot, 1996] tries to reduce the amount of information in the system to the basic semantics of the analysed object to be able to give answers to questions which would be undecidable otherways. For example a tool can easily detect type mismatches by abstracting from the actual value of a variable to its type.

**Clone Detection** [van Rysselberghe et al., 2004] searches for repeated occurrences of similar structures, marking these occurrences as clones. Such clones do not only increase the amount of information to be managed, but they also increase the chance that when changing one occurrence, other occurrences are overlooked.

**Pattern matching** [Beck et. al., 1999] or smell detection is the most popular approach in static analysis. For high-level programming languages like COBOL, C, C# or Java commonly used in traditional software development, feasible patterns have been empirically proven since decades. Just like a virus scanner, the tool tries to find areas which are similar to pre-defined patterns ("code smells") and tags them with a probability representing the tool's confidence in the finding.

**4. STUDY OF SPREADSHEET AUDITING TOOLS**

To approach the questions mentioned at the beginning, the first author has set up a small student research project [Berberich et. al., 2012] in an industrial context. Prior to the study, we had already established contact with a department of an industry partner (we will refer it to as DEPT) which employs around 150 technicians who work in the field of measuring and controlling emissions of industry plants. In this process, they use a dozen spreadsheet templates to capture the emission values, compute violations of allowed emission levels and produce reports about them.

The DEPT employees have used their spreadsheet templates for more than 10 years and have produced around 1500 concrete spreadsheets during this time. The templates are well-established and modifications to single templates do not occur more than twice per quarter. DEPT has an internal quality assurance process which prescribes that each single report sent out to external parties has to undergo an inspection according to the four-eyes-principle. Although DEPT have not struggled often with severe spreadsheet errors in their reports, they find it just time-consuming to inspect their spreadsheets. Their main problem was that they applied the technique "carefully inspect each cell" without any dedicated tool support and without knowing what to look for in particular.

| Sample | # of Cells | # of Worksheets | # of Formulas |
|--------|------------|-----------------|---------------|
| 1 | 2714 | 10 | 178 |
| 2 | 3236 | 10 | 184 |
| 3 | 3236 | 10 | 186 |
| 4 | 610 | 2 | 117 |

| 5 | 2151 | 6 | 182 |
| 6 | 2990 | 4 | 208 |
| 7 | 2092 | 4 | 119 |
| 8 | 2356 | 4 | 225 |
| 9 | 1080 | 3 | 75 |
| 10 | 2034 | 1 | 58 |
| 11 | 6625 | 7 | 2432 |

*Figure 1: Basic metrics of the spreadsheets supplied by our industry partner*

To help the industry partner in this situation, three undergraduate Software Engineering students were assigned the task to find spreadsheet auditing tools suitable for the needs of DEPT in a student research project scheduled for 4 months. As their first step, the students precisely analyzed the requirements of the industry partner and scanned the market for available spreadsheet auditing tools. They found a total of 14 tools ranging from open source research prototypes to aggressively marketed commercial products. This included tools implemented as spreadsheets themselves, Add-Ins for Microsoft Excel, standalone tools and "scanning services" in the form of SaaS (Software as a Service) on the web. For each of the tools we tried to get an evaluation license to give the students the chance to actually run the tools. Although most vendors were very interested in this study, a few vendors did not reply to our evaluation request or offered their tools only under unacceptable terms, resulting in these tools being removed from the selection. Also, the students developed a list of knockout criteria to shorten the list of tools to be analyzed in detail.

Only three tools made it to the shortlist. The students thoroughly tried them out on 11 concrete spreadsheets supplied by our industry partner (see Figure 1) and evaluated them against eight functional and non-functional requirements gathered earlier from the analysis. Then, the students presented the tools to the industry partner to check if their own findings also matched the perceptions of the industry partner. At the end, the students were able to give a clear recommendation for one of the three tools. Because one of the industry partner's most limiting knockout criteria were the licensing costs, this was rather a "value for money" recommendation than one for the "best" tool available on the market at the time of the study.

**5. PRACTICAL CHALLENGES**

Now, three months after completion of this study, we have reflected on its results and experiences with all 14 tools involved. In the following, we share our impressions that are purely based on this reflection. Our aim is not to give concrete recommendations for specific tools but to stress widespread issues and foster a discussion about possible solution approaches.

We have the impression that most of the 14 tools do a satisfactory job in scanning spreadsheets for smell patterns like constants in formulas, references to blank cells or complex formulas. Unlike many tools for software engineers, most spreadsheet auditing tools require only small efforts for installation and the first automated inspection can be initiated easily without the help of specialists.

Many auditing features are already built into modern spreadsheet execution environments, but these features are hidden or hardly accessible for casual users. A good example is the powerful "goto -> special" functionality in Microsoft Excel. Many spreadsheet auditing tools enhance these features, i.e. by comprehensive lists of error conditions like "DIV/0" or discovering unprotected formulas.

However, we have identified several areas where we see practical challenges, since (nearly) all of the reviewed tools had significant problems providing a satisfactory experience for their users in these areas. These areas are presented in this section, ordered by our subjective perception of their severity:

**5.1 Presentation of findings**

The primary output of a spreadsheet auditing tool which employs fault-detection patterns is a list of findings in the spreadsheets it was instructed to scan. It is essential for this presentation to be useful and understandable to the user. Therefore, the presentation must issue a proper abstraction level (i.e. by grouping findings) and allow users to easily trace each particular finding down to the suspicious cells.

Most spreadsheet auditing tools on the market tend to produce flat lists with findings, referring to various cells. Especially when the findings are output in separate documents the navigation from the finding to the affected cell is often cumbersome. Only few tools allow users to automatically navigate to the finding without locating the cells manually. To make it worse, in many cases such lists are not ordered (i.e. by severity of the findings) or grouped (i.e. by type of the finding).
Many tools color cells or add comments to a spreadsheet, as shown exemplary in Figure 2. We experienced several problems with the approach of coloring cells:

- The "flagging" colors are rarely customizable.
- Colors which are used to show severities (i.e. red=bad, green=good) might seem intuitive, but often colors are used to represent similarities and differences. Here, they often implicitly transport hidden messages like "dark brown is more similar to red than to blue" which are not intended.
- We have seen tools which used more than 32 colors, which is way too much because humans have huge problems in distinguishing them. For instance, if the tool uses "light blue", "navy blue" and "medium blue" across a huge spreadsheet with several worksheets, this presentation is mostly useless.

*Figure 2: Excerpt of a spreadsheet before (a) and after (b) coloring by an auditing tool*

In general, the use of both colors and comments for communicating findings to the user is problematic. Colors and comments are often already used as part of the original spreadsheet. Auditing tools try to address this issue by removing the original colors and comments before running the inspection. Unfortunately, there are spreadsheets which heavily rely on conditional formatting and transport their main information through colors, not numbers. Having them removed or replaced can lead to the effect of spreadsheet users not recognizing their own spreadsheets anymore and, thus, being unable to trace the findings.

Apart from that, many tools only provide a presentation which is useful for "going through each finding" but they lack a management summary that puts the number and severity of findings in relation to other metrics like the size of the spreadsheet. Some of the "control & compliance" tools do a better job here, but the views they provide often lack navigation functions which are important for traceability.

In static analysis for traditional software development there is also a lot of information (fault area, pattern matched, estimated severity) to be presented to the user, which makes an optimal presentation difficult. Commercial vendors of TSATs have invested much effort to increase the acceptance of their reporting because they realized that only a tool which provides understandable results will sell.

**5.2 Handling of false positives**

Spreadsheet auditing tools are known to report high numbers of false positives. Researchers report rates between 78% [Cunha et. al., 2012] and 90% [Powell et. al., 2008]. Because the tools don't work with specifications provided by the users, they have to make their own assumptions which can be wrong. Therefore, false positives are unavoidable. As a consequence, it is important that tools provide proper means to handle false positives. Going through a list of findings and flagging false positives is a tedious and time-consuming work, and users should not have to do it more than once - unless new findings appear. Unfortunately, almost all spreadsheet auditing tools don't provide any means to this end - on subsequent runs, the user would get the same list of findings, including all findings he or she already flagged as false positives.

The handling of false positives is still an area to be improved in TSATs as well, although TSATs do not produce such high numbers of false positives. Usually, a thorough configuration of the tools helps reducing the amount, but help for marking them and keeping track of them over time is only inadequately implemented in most tools by providing blacklists or working with annotations within the analysed code. These approaches are not perfect and not accepted enthusiastically by the community, but they are still better than the complete lack of support for handling false positives as we observed it in most spreadsheet auditing tools.

**5.3 Non-intrusiveness**

Many spreadsheet auditing tools alter the original spreadsheet to be inspected during the inspection to render their reports in one of the following ways:

- Cells are colored, indicating "high risk areas" or structural similarities and differences.
- Comments are added to cells.
- Additional worksheets with reports or colored versions of the original worksheets are appended.
- The spreadsheet is enriched by parts of the tool itself (i.e. VBA functions).

Such use can lead to unwanted side-effects, especially when copied cells contain formulas referring to the original sheets. In extreme cases, this could even result in spreadsheets computing wrong values not due to faults in the spreadsheets themselves but due to faults injected by such spreadsheet auditing tools. Some tools try to remedy the problem by creating copies of the complete spreadsheet but this does not provide enough uncoupling in case formulas in these spreadsheets refer to other spreadsheets.

In TSATs this problem does not occur because all additional information is shown on an IDE overlay and is not included in the analysis object. Especially for safety-related software, the risky approach taken by many spreadsheet auditing tools is considered out of the question.

**5.4 Understandability of implemented rules**

Most spreadsheet auditing tools report matches with smell-patterns as "errors". But in many cases, end-users see cells or formulas which they regard as being correct flagged as errors and often they don't understand what the tool is trying to tell them about these cells. In fact, many of the smells are rather related to "higher qualities" like maintainability, but many tools we have seen failed to provide enough hints and explanations to justify their criticism.

TSATs have exactly the same problem. Research in this area such as [Bessey et al., 2010] shows that highly sophisticated analyses do not increase the value of a tool for the user because users tend to ignore warnings or dismiss them as false positives if they do not understand why they were raised.

**5.5 Configurability and result inter-comparability**

As Zitzelsberger has demonstrated [Zitzelsberger, 2012], many spreadsheet auditing tools report very different numbers of findings even for very simple patterns like "constants in formulas". The reason for this is that many tools make hidden assumptions and exceptions, i.e. by not treating the numbers 0 and 1 as constants in formulas. In many cases, this behavior is neither configurable nor transparent. The tools only allow users to completely disable particular checks.

In contrast, most of the TSATs are highly configurable and allow to precisely define exceptions and configure the internal thresholds of their scanners. This makes them more understandable, more comparable and also helps to reduce false positives.

**5.6 Lack of unification**

As previously discussed, spreadsheet auditing tools try to help users with many different techniques. Therefore, it is not surprising that there is no "Swiss army knife" tool which performs great in all disciplines. Instead, specific tools are useful for specific purposes. We do not expect this situation to change in the near future. But for end-users who want to audit their spreadsheets this means that they have to deal with completely different interfaces for configuring and running the tools, as well as with completely different ways to receive the results.

Because of the many ways to detect potential problems using static analysis, there is also no TSAT which detects everything. But here, this problem is conquered by creating so-called dashboards which aggregate the results of different tools. Figure 3 shows an example for such a dashboard (taken from "conQAT" [Deissenboeck et. al., 2008]).

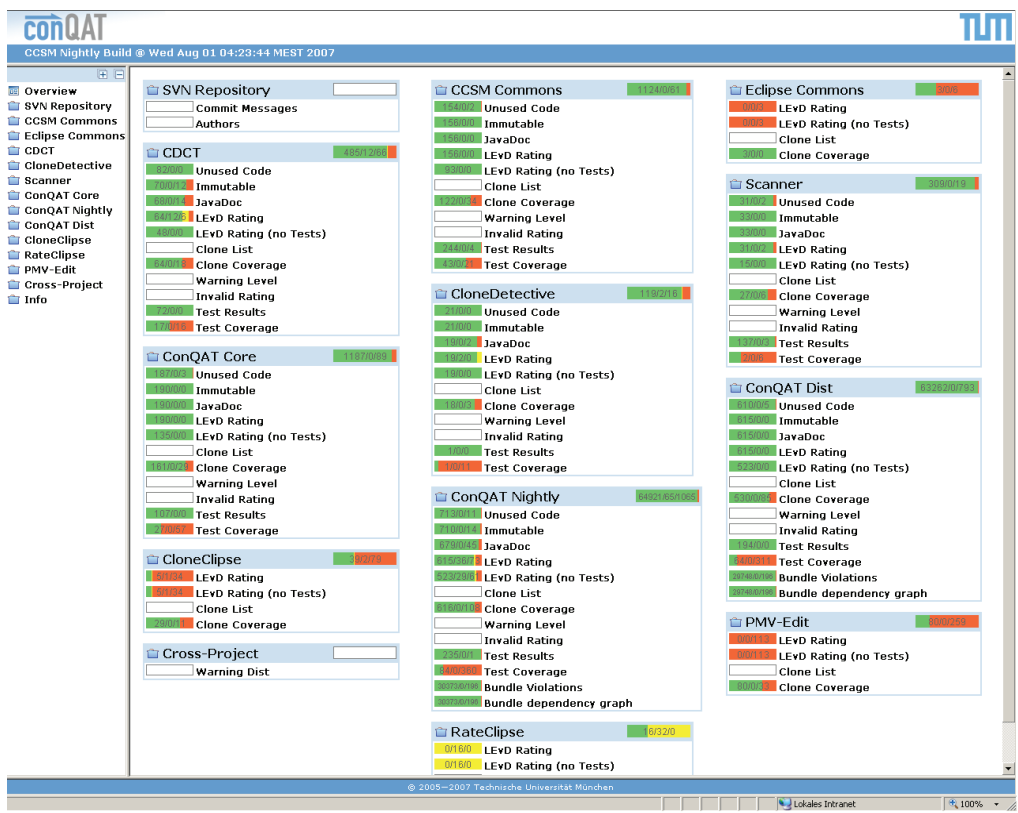

*Figure 3: Dashboard aggregating result from several tools, generated by conQAT*

**5.7 Licensing costs versus risk perception**

Our industry partner did not perceive faulty spreadsheets as a major risk. Although this might be explained by the phenomenon of "overconfidence" [Panko 2007], the claims in literature could not stand up to the practical experience of the industry partner having seen few and only insignificant errors in years of spreadsheet use. Therefore, DEPT sees spreadsheet auditing tools rather as a "nice-to-have" and not as a "must-have" and is not willing to spend more than 1000€ for five licenses of such a tool. Most spreadsheet auditing tools cost much more than that. For us, the only explanation for this is that these tools do not target "yet spreadsheet-risk unconvinced" organizations but rather "convinced" organizations or those that are forced (i.e. by law) to audit their spreadsheets.

The question of the cost for risk reduction is of course also an issue for traditional software development, especially in areas like software for medical devices. There is often more at stake and lower overconfidence, so the developers use every help they can get. Furthermore, there are binding standards like the [MISRA] in the automotive industry which forces the developers to use a tool to show their compliance with the standard, while for spreadsheets there are mostly just proposals like [FAST].

**5.8 Compatibility and Portability**

Many spreadsheet auditing tools are tied closely to particular spreadsheet execution environments. They often realize their internal core functionality by using the APIs these execution environments provide. While this allows the tools to audit every aspect of a spreadsheet compatible with the particular execution environment, it also makes the tools less portable. As these APIs often change between major releases of spreadsheet execution environments, tool vendors are forced to issue major updates to their tools. For instance, there was a major API change between Microsoft Excel 2003 and 2007, leading to a situation where many spreadsheet auditing tools which were compatible with Excel 2003 had to be rewritten from scratch in order to provide support for Excel 2007 as well. And tools developed for Excel 2007 and above often do not provide backwards compatibility with Excel 2003 and below.

In our opinion, tool vendors should strive to develop the core components of their analysis tools decoupled from particular execution environments. They should audit the spreadsheets based on a data model of their own which was designed for the inspection of spreadsheets from the beginning and is only loosely coupled with the spreadsheets' internal data format by using independent libraries. Therefore, the auditing core should be fully functional without the execution environment. The latter should only be used to provide a thin user interface which connects the functionality of the auditing core with the spreadsheet execution environment. As far as we know, only spreadsheet auditing tools originating from research like "Smellsheet Detective" [Cunha et. al, 2012] or "Spreadsheet Inspection Framework" [Zitzelsberger, 2012] adopt these principles.

TSATs do not run into similar problems because they only rely on formal definitions of high level programming languages. This makes them independent of APIs which are known to change more frequently than programming languages. This can be easily adopted to spreadsheets because the basic concepts of having cells connected by formulas did not significantly change since VisiCalc introduced it.

**5.9 Localization**

DEPT's IT department supplies software to its employees. All software is provided with a German user interface. Therefore, DEPT's end-users expect to get any new tool - including a spreadsheet auditing tool - with a German user interface. In general, most spreadsheet users at DEPT speak and understand English well, but many of them don't know the English spreadsheet terminology and, thus, don't understand findings like "spurious reference".

Surprisingly, none of the tools evaluated in the study provided any localization for German environments.

**6 CONCLUSION**

We have seen that spreadsheet auditing tools share many commonalities with static analysis tools for traditional software development. As the spreadsheet auditing tools are relatively young, they have not yet reached the maturity of the tools used in traditional software development. Using spreadsheet auditing tools seems to be more efficient than doing pure manual inspections, but this efficiency could be greatly enhanced if their vendors would apply the lessons learned from the tools used in traditional software development. The latter ones have already developed working solutions in areas where spreadsheet auditing tools are still having a hard time although these solutions appear to be directly transferable.

**ACKNOWLEDGEMENTS**

This study on which this work reflects would have not been possible without the help of our three students and our industry partner. We would also like to thank all tool vendors who provided us with evaluation versions of their tools. We are also very grateful to Maarten Bessems from spreadsheetsoftware.com who has helped us a lot with his insight and his expertise on both available tools and common issues he experienced in past years when auditing complex spreadsheets in industry. Last but not least, we have to thank Jochen Ludewig, Kornelia Kuhle, Eugen Massini and the two anonymous reviewers for their helpful remarks regarding earlier versions of this paper.

# REFERENCES

Beck, K., Fowler, M., & Beck, G. (1999), "Bad smells in code. Refactoring: Improving the design of existing code", pages 75-88.

Berberich, T., Nguyen, A. and Vetter, M. (2012), "Spreadsheet Auditing Tools", Case Study, University of Stuttgart

Bessey, A., Block, K., Chelf, B., Chou, A., Fulton, B., Hallem, S., ... & Engler, D. (2010), "A few billion lines of code later: using static analysis to find bugs in the real world", Communications of the ACM, 53(2), pages 66-75

Clermont, M., Hanin, C., Mittermeir, R. (2008), "A Spreadsheet Auditing Tool Evaluated in an Industrial Context", Proceedings of the 2002 EuSpRIG Conference

Cousot, P. (1996), "Abstract interpretation", ACM Comput. Surv. 28, 2 (June 1996), pages 324-328.

Cunha, J., Fernandes, J. P., Martins, P., Mendes, J., & Saraiva, J. (2012), "Smellsheet detective: A tool for detecting bad smells in spreadsheets", Proceedings of the 2012 IEEE Symposium on Visual Languages and Human-Centric Computing. (Vol. 12).

Cunha, J., Fernandes,J.P., Peixoto, C., Saraiva, J. (2012), "A Quality Model for Spreadsheets", Proceedings of the 8th International Conference on the Quality of Information and Communications Technology, pages 231-236

Deissenboeck, F., Juergens, E., Hummel, B., Wagner, S., Mas y Parareda, B., & Pizka, M. (2008), "Tool support for continuous quality control. Software", IEEE,25(5), pages 60-67.

FAST: "The FAST Modelling Standard", http://www.fast-standard.org  04.30pm  02/28/2013

Hermans, F. (2013), "Analyzing and visualizing spreadsheets", PhD thesis, TU Delft

Howard, P. (2005), "Managing Spreadsheets", White Paper. Bloor Research, Suite, 4.

Kulesz, D. (2011), "From Good Practices to Effective Policies for Preventing Errors in Spreadsheets", Proceedings of the EuSpRIG 2011 conference

Mäntylä, M. (2003), "Bad smells in software-a taxonomy and an empirical study", Helsinki University of Technology

MISRA: http://www.misra.org.uk 04.30pm  02/28/2013

Panko, R. R. (2007), "Two experiments in reducing overconfidence in spreadsheet development", Journal of Organizational and End User Computing (JOEUC), 19(1), pages 1-23

Powell, S. G., Baker, K. R., & Lawson, B. (2008), "An auditing protocol for spreadsheet models. Information & Management" 45(5), pages 312-320

Taylor, R. N., & Osterweil, L. J. (1980), "Anomaly detection in concurrent software by static data flow analysis", IEEE Transactions on Software Engineering, pages 265-278

Van Rysselberghe, F., & Demeyer, S. (2004), "Evaluating clone detection techniques from a refactoring perspective", Proceedings. 19th International Conference on Automated Software Engineering, pages 336-339

Zitzelsberger, S. (2012), "Error Detection in Spreadsheets", Diploma Thesis, University of  Stuttgart